\documentclass[runningheads]{svmult}

\usepackage{makeidx}   
\usepackage{graphicx}  
\usepackage{subeqnar}  
\usepackage{multicol}  
\usepackage{physprbb}  
\makeindex             

\begin{document}
\title*{Does rotation of B stars depend on metallicity?\protect\newline
preliminary results from GIRAFFE spectra}
\toctitle{Does rotation of B stars depend on metallicity?
\protect\newline preliminary results from GIRAFFE spectra}
%
%
\titlerunning{Does rotation of B stars depend on metallicity?}
%
\author{Fr\'ed\'eric Royer\inst{1}
\and Pierre North\inst{2}
\and Claudio Melo\inst{3}
\and Jean-Claude Mermilliod\inst{2}
\and Eva~K.~Grebel\inst{4}
\and Jose Renan de Medeiros\inst{5}
\and Andr\'e Maeder\inst{6}}
\authorrunning{Fr\'ed\'eric Royer et al.}
%
%
\institute{GEPI / CNRS UMR 8111, Observatoire de Paris, F-92195 MEUDON cedex,
France
\and Laboratoire d'Astrophysique, Ecole Polytechnique F\'ed\'erale
de Lausanne (EPFL), Observatoire,
CH-1290~Sauverny, Switzerland
\and ESO, Alonso de Cordova 3107, Casilla 19001, Santiago 19, Chile
\and Astronomisches Institut, Universit\"at Basel, Venusstr. 7,
CH-4102 Binningen, Switzerland
\and Universidade Federal do Rio Grande do Norte, 59072-970 Natal, R.N., Brazil
\and Observatoire de Gen\`eve, CH-1290 Sauverny, Switzerland}

\maketitle              

\begin{abstract}
We show the $v\sin i$ distribution of main sequence B stars in sites of
various metallicities, in the absolute magnitude range $-3.34 < M_V < -2.17$.
These include Galactic stars in the field measured by \cite{nor:ALG02}, members
of the h \& $\chi$ Per open clusters measured by \cite{nor:N04}, and
five fields in the SMC and LMC measured at ESO Paranal with the
FLAMES-GIRAFFE spectrograph, within the Geneva-Lausanne guaranteed time.
Following the suggestion by \cite{nor:MGM99},
we do find a higher rate of rapid rotators in the Magellanic Clouds than in 
the Galaxy, but the $v\sin i$ distribution is the same in the LMC and in the
SMC in spite of their very different metallicities.
\end{abstract}

\section{Introduction, results and conclusion}
This work aims at testing the suggestion of \cite{nor:MGM99}
that stellar rotation is faster at lower metallicity by direct measurements,
especially in the LMC and SMC, on stars with $-3.34 < M_V < -2.17$, i.e.
spectral types B0-B6 or masses from $\sim 6.7$ to 14 M$_\odot$. This work is
complementary to that of \cite{nor:K04}, which deals with slightly more massive
stars. The results are shown on Fig.~1 and commented in the caption.
There is an excess of slow rotators in the Galaxy relative to the MCs,
but the $v\sin i$ distributions of the LMC and the SMC are surprisingly
similar.

\begin{figure}[b]
\begin{center}
\includegraphics[width=1\textwidth]{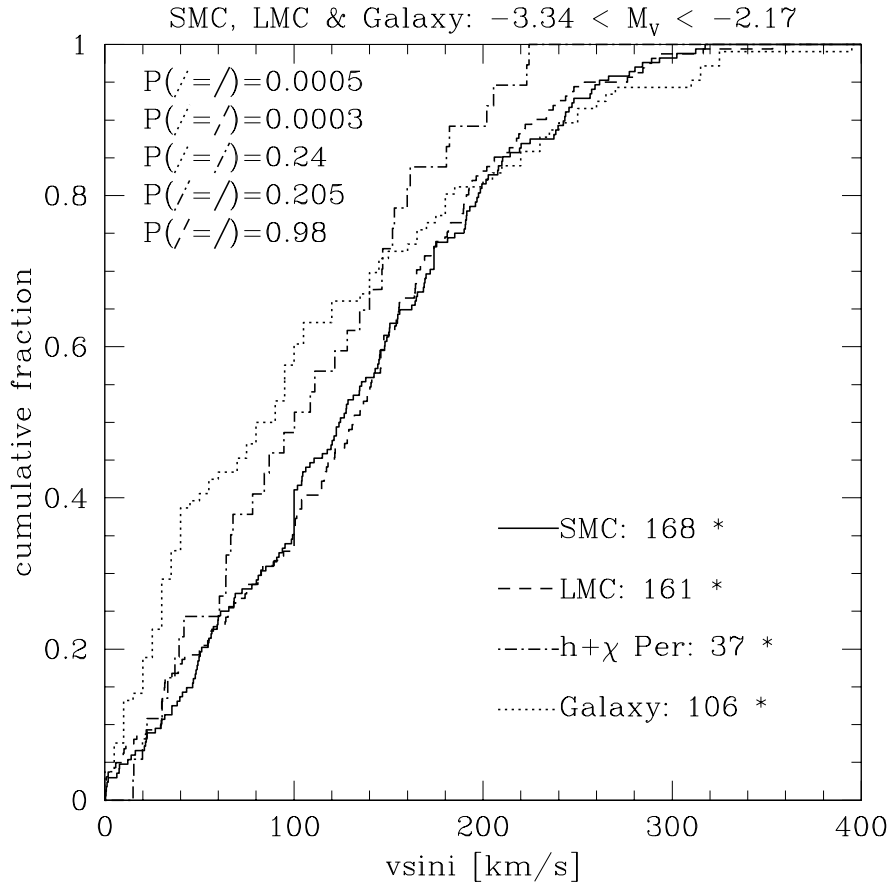}
\end{center}
\caption[]{Cumulative $v\sin i$ distributions for Galactic stars in the field
(dotted line), for members of the h \& $\chi$ Per clusters (dash-dot) and for
stars in the LMC (short dash) and in the SMC (solid line).
The GIRAFFE spectrograph, attached to the UT2 telescope (VLT) and used
in the L2 setup ($R=6400$, $\lambda_c=4272$~\AA), was used on 3 fields in the
LMC (centered on
[$\alpha_{J2000}=$05:31:40,$\delta_{J2000}=$-66:59:48], on [05:30:40,
-67:17:12] and on [05:03:48, -69:00:36]) and 2 fields in the SMC (centered on
[00:56:12, -72:29:00] and on [00:49:26, -73:12:07]).
 We fitted synthetic spectra to observed ones in the range
$4460-4490$~\AA~ with the technique described by \cite{nor:ENa}
using an average $T_{\rm eff}-M_V$ relation for the main sequence and assuming
$\log g=4.0$. The resulting $v\sin i$ values were then transformed to
the scale of \cite{nor:S75}.
For the Galaxy, we defined the $v\sin i$ distribution using 1) the measurements
made in the h \& $\chi$ Per clusters by \cite{nor:N04} and
2) the large sample of \cite{nor:ALG02} of bright field B stars, Geneva
photometry being used to determine $M_V$ through the calibration of \cite{nor:C99}.
The SB2 systems were eliminated from this sample, which, although
magnitude-limited, does not significantly differ from a volume limited one.
The results are summarized in this Figure. Surprisingly, the overall
$v\sin i$ distribution is almost exactly the same in the SMC (mean metallicity
$Z\sim 0.008$) and in the LMC ($Z\sim 0.004$). There is only
a marginal difference between h \& $\chi$ Per ($Z(\textrm{h Per})\sim 0.01$
according to \cite{nor:SMS04}) and the MCs, but a very significant
one ($P<0.1$~\%) between the Galactic field ($Z\sim Z_\odot=0.018$) and the
MC fields. Thus, either the
metallicity effect saturates for $Z < Z(\textrm{LMC})\sim 0.008$, or another
cause affects rotational velocities, e.g. different rates and orbital
parameters of SB1 binaries (not excluded from the samples), through tidal
effects.
}
\end{figure}

%

\end{document}